\documentclass[aps,prl,twocolumn,superscriptaddress]{revtex4}
\usepackage{graphicx}
\begin{document}
\title{Evidence for ferromagnetic spin-pairing superconductivity in UGe$_2$: \\
A $^{73}$Ge-NQR study under pressure}

\author{A.~Harada}\email[aharada@nmr.mp.es.osaka-u.ac.jp]{}
\affiliation{Department of Materials Engineering Science, Osaka University, Osaka 560-8531, Japan}
\author{S.~Kawasaki}
\affiliation{Department of Materials Engineering Science, Osaka University, Osaka 560-8531, Japan}
\author{H.~Mukuda}
\affiliation{Department of Materials Engineering Science, Osaka University, Osaka 560-8531, Japan}
\author{Y.~Kitaoka}
\affiliation{Department of Materials Engineering Science, Osaka University, Osaka 560-8531, Japan}
\author{Y.~Haga}
\affiliation{Advanced Science Research Center, Japan Atomic Energy Research Institute, Tokai, Ibaraki 319-1195, Japan}
\author{E.~Yamamoto}
\affiliation{Advanced Science Research Center, Japan Atomic Energy Research Institute, Tokai, Ibaraki 319-1195, Japan}
\author{Y.~\={O}nuki}
\affiliation{Advanced Science Research Center, Japan Atomic Energy Research Institute, Tokai, Ibaraki 319-1195, Japan}
\affiliation{Department of Physics, Osaka University, Osaka 560-0043, Japan}
\author{K.~M.~Itoh}
\affiliation{Department of Applied Physics and Physico-Informatics, Keio University, Yokohama 223-8522, Japan}
\author{E.~E.~Haller}
\affiliation{University of California at Berkeley and Lawrence Berkeley National Laboratory, Berkeley, CA 94720, USA}
\author{H.~Harima}
\affiliation{Department of Physics, Faculty of Science, Kobe University, Nada, Kobe 657-8501, Japan}

\date{\today}

\begin{abstract}
We report that a novel type of superconducting order parameter has been realized in the ferromagnetic states in UGe$_2$ via $^{73}$Ge nuclear-quadrupole-resonance (NQR) experiments performed under pressure ($P$). Measurements of the nuclear spin-lattice relaxation rate $(1/T_1)$ have revealed an unconventional nature of superconductivity such that the up-spin band is gapped with line nodes, but the down-spin band remains gapless at the Fermi level.  This result is consistent with that of a ferromagnetic spin-pairing model in which Cooper pairs are formed among ferromagnetically polarized electrons.  The present experiment has shed new light on a possible origin of ferromagnetic superconductivity, which is mediated by ferromagnetic spin-density fluctuations relevant to the first-order transition inside the ferromagnetic states.
\end{abstract}
\vspace*{5mm}
\maketitle
The coexistence of magnetism and superconductivity (SC) has recently become an important topic in condensed-matter physics. The recent discovery of SC in ferromagnets UGe$_2$ \cite{Saxena,Huxley} and URhGe \cite{Aoki} has been a great surprise because the Cooper pairs are influenced by a non-vanishing internal field due to the onset of ferromagnetism (FM), which is believed to prevent the onset of SC. In the ferromagnet UGe$_2$ with a Curie temperature $T_{\rm Curie}=52$\,K at ambient pressure ($P=0$), the emergence of $P$-induced SC has observed in the $P$ range of 1.0\,-\,1.6\,GPa \cite{Saxena,Huxley}. It is noteworthy that the SC in UGe$_2$ disappears above $P_{\rm c}\sim 1.6$\,GPa, beyond which FM is suppressed. The SC and FM in this compound have been shown to be cooperative phenomena \cite{Harada}. The superconducting transition temperature ($T_{\rm sc}$) is the highest at $P_{\rm x}\sim 1.2$\,GPa, where a first-order transition occurs from FM2 to FM1 as $P$ increases. Here, it should be noted that ferromagnetic moments are increased in the first-order transition from FM1 to FM2 as functions of temperature and pressure, as shown in Fig.~1(a) \cite{Pfleiderer,Tateiwa1,Kotegawa}. The $P$-induced SC in UGe$_2$ coexists with FM1 and FM2 exhibiting the large magnetization of an order 1\,$\mu_{\rm B}$/U even for the case of $T_{\rm Curie}\sim$ 30\,K \cite{Kotegawa}. Therefore, the onset of SC is proposed to be suitable for the formation of a spin-triplet pairing state rather than a spin-singlet pairing state \cite{Huxley}. However, there are few reports that address the type of order parameter is realized in FM1 and FM2. 
%fig1
\begin{figure}[h]
\centering
\includegraphics[width=7.5cm]{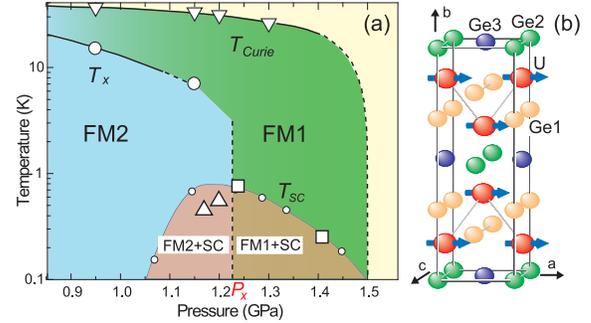}
\caption[]{(color online). \footnotesize (a) Pressure versus temperature phase diagram of UGe$_2$ near the superconducting phase \cite{Kotegawa}. The $T_{\rm sc}$ values of FM1 (open squares) and FM2 (open triangles), and $P_{\rm x}$ value determined in this study are plotted. (b) Crystal structure of UGe$_2$ with the ferromagnetic moment at the U site below $T_{\rm Curie}$.}
\end{figure}  
In a previous study, an unconventional nature of the SC has been suggested from the measurement of the $^{73}$Ge-NQR nuclear spin-lattice relaxation rate $1/T_1$ \cite{Kotegawa}.  However, it has not been well understood whether the presence of the residual density of states (RDOS) at the Fermi level in the SC state is intrinsic or not, suggesting the occurrence of a possible extrinsic effect due to the presence of any impurity and/or imperfection in the sample \cite{Kotegawa}.  In particular, it is unclear why SC emerges with the highest $T_{\rm sc}$ when the first-order transition occurs from FM1 to FM2 at $P_{\rm x}\sim 1.2$\,GPa.  In order to gain insight into this issue, further experiments are required for understanding a $P$-induced evolution in the FM states and a novel order-parameter symmetry emerging in the FM states in UGe$_2$.
 
In this letter, by performing the $^{73}$Ge-NQR measurements under pressure at zero field ($H=0$) on a newly prepared sample, we report that the SC in this compound is caused by the formation of up-spin Cooper pairs, where the gap opens only at the up-spin band in FM1 and FM2 but not at the down-spin band \cite{Ohmi,Machida,Fomin}. 
The ferromagnetic spin-pairing SC is considered to be mediated by ferromagnetic spin-density fluctuations relevant to the first-order transition inside the ferromagnetic states.
%fig1
\begin{figure}[h]
\centering
\includegraphics[width=7.8cm]{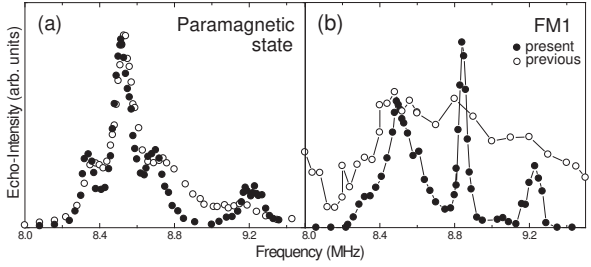}
\caption[]{(color online). \footnotesize Comparison of the $^{73}$Ge-NQR spectra of the present and previous samples of UGe$_2$ in (a) paramagnetic state and (b) FM1. The spectra at (a) $P=1.9$\,GPa and (b) $P=$ 1.41 and 1.3\,GPa are shown by solid and open circles, respectively, demonstrating that the present sample has better quality than the previous samples \cite{Kotegawa}.}
\end{figure}

%\section{Experimental}
A polycrystalline sample enriched by $^{73}$Ge  was crushed into coarse powder for the NQR measurement and annealed to maintain its quality.  The NQR experiments were performed by the conventional spin-echo method at $H=0$ in the frequency ($f$) range of 5\,-\,11\,MHz at $P$ = 1.17, 1.2, 1.24, and 1.41\,GPa. Hydrostatic pressure was applied by utilizing a NiCrAl-BeCu piston-cylinder cell filled with Daphne oil (7373) as a pressure-transmitting medium. The value of $P$ at low temperatures was determined from the $T_{\rm sc}$ of Sn measured by the resistivity measurement.  The possible distribution of the pressure inside the sample was less than 3\,$\%$ in the present experimental setup. A $^{3}$He-$^{4}$He dilution refrigerator was used to obtain the lowest temperature of 50\,mK. Figures~2(a) and 2(b) show the NQR spectra in paramagnetic (PM) phase and FM1 phase, respectively. The linewidths in these NQR spectra are narrower for the present sample than for the previous sample, demonstrating that the quality of the present sample is significantly higher than that of the previous sample. Moreover, the NQR-$T_1$ measurements reveal that the present sample exhibits the highest value of $T_{\rm sc}=0.75$\,K obtained thus far at $P=1.24$\,GPa, ensuring higher quality than before.  
%%fig2
\begin{figure}[h]
\centering
\includegraphics[width=6.5cm]{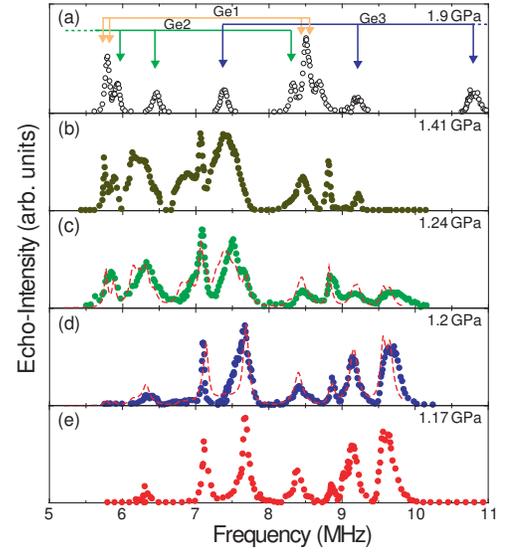}
\caption[]{(color online). \footnotesize (a) $^{73}$Ge-NQR spectra at 4.2\,K in the $P$-induced paramagnetic phase. 
The NQR spectra in (b), (c), (d), and (e) represent in the ferromagnetic phases at 1.4\,K and $P=$ 1.41, 1.24, 1.2 and 1.17\,GPa, respectively. The dashed lines in the figures indicate the simulated results (see text). }
\end{figure} 

Figure~3(a) shows the NQR spectra for the PM phase at 4.2\,K and $P=1.9$\,GPa where FM1 is completely suppressed. They reveal a structure consisting of separated peaks associated with three inequivalent Ge sites in one unit cell in the crystal structure illustrated in Fig.~1(b) \cite{Harada,Kotegawa}. The number of Ge1 sites is twice that of Ge2 and Ge3 sites in one unit cell. The Ge1 site is closely located along the uranium (U)-zigzag chain, while the other two sites Ge2 and Ge3 are located outside this zigzag chain.  From the analysis of the NQR spectra for FM1 in the previous experiment \cite{Kotegawa}, it was demonstrated that the onset of FM1 induces an internal field $H_{\rm int}=0.9$\,T at the Ge sites that additionally causes about the Zeeman splitting in each Ge-NQR spectrum. Furthermore, the angle between the principal axis for the nuclear quadrupole Hamiltonian and a direction of $H_{\rm int}$ was determined as $\theta\sim\pi/3$.  When the first-order transition occurs from FM1 to FM2, the spectral shape changes significantly from the spectra at $P=1.24$ and 1.41\,GPa to those at $P=1.17$ and 1.2\,GPa, as shown in Figs.~3(b)-(e). From the analysis of the spectra, it is estimated that $H_{\rm int}=0.9$\,T for FM1 increases to $H_{\rm int}=1.8$\,T for FM2 (see Fig.~5(b)). The sudden increase in $H_{\rm int}$ in the narrow $P$ range should be relevant to the first-order transition at $P_{\rm x}$.  In such a case, the spectra near $P_{\rm x}$ are expected to reveal a mixture of both domains of FM1 and FM2 in the narrow range of $P$ close to $P_{\rm x}$ due to an inevitable distribution of $P$.
In fact, the respective spectra at $P=1.2$ and 1.24\,GPa are composed of spectra arising from FM1 ($P=1.41$\,GPa) and FM2 ($P=1.17$\,GPa) with the ratios of 1\,:\,9 and 7\,:\,3, respectively, as shown by the dashed lines in Figs.~3(b) and 3(c). By considering an inevitable $P$ distribution ($\Delta P = 0.04$\,GPa) as a Gaussian function given by $\exp[- ([P-P_0]/[\Delta P/(2 \sqrt{\ln 2})])^2 ]$, we obtain $P_{\rm x}=1.23$\,GPa.
The present experimental results reveal that the first-order transition occurs at $P_{\rm x}=1.23$\,GPa. 
%%fig3
\begin{figure}[hb]
\centering
\includegraphics[width=6.8cm]{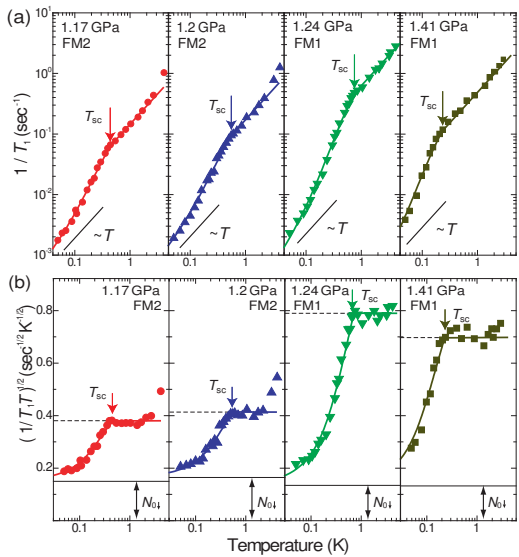}
\caption[]{(color online). \footnotesize (a) Temperature dependences of $1/T_1$ for FM2 at $P=1.17$ and 1.2\,GPa measured at $f=7.68$\,MHz and for FM1 at $P=1.24$ and 1.41\,GPa measured at 7.09 and 7.07\,MHz, respectively. (b) Temperature dependences of $(1/T_1T)^{1/2}$ related to the DOS in either the SC state or the normal state at each $P$. 
The solid curves represent the results calculated based on the ferromagnetic spin-pairing model (see the text).}
\end{figure} 
\\
\indent
Figure~4(a) shows the $T$ dependences of $1/T_1$ for FM1 and FM2 at pressures that are slightly lower and higher than $P_{\rm x}=1.23$\,GPa, respectively. Clearly, $1/T_1$ for FM1 and FM2 decreases without any indication of coherence peak just below $T_{\rm sc}$, which provides evidence for the uniform coexistence of the unconventional SC and ferromagnetism. In the previous study \cite{Kotegawa}, the line-node gap model with RDOS $N_{\rm res}$ at the Fermi level was applied to interpret a systematic evolution in the superconducting energy gap $\Delta$ and a fraction of RDOS $N_{\rm res}$/$N_{0}$.  Here $N_0$ is the density of state (DOS) at the Fermi level in the FM phases. It should be noted that all the data of $1/T_1$ are uniquely determined in the present sample, but not in the previous sample \cite{Kotegawa}. Therefore, we could not exclude the fact that the RDOS is present in the previous sample due to some impurity effect. Similarly for the present sample with higher quality than that of the previous sample, the application of the line-node gap model with the RDOS allows us to estimate $N_{\rm res}$/$N_{0}$~= 0.50, 0.48, 0.29, and 0.30 and $T_{\rm sc}=$ 0.45, 0.55, 0.75, and 0.25\,K ($\pm0.05$\,K) at $P=$ 1.17, 1.2, 1.24, and 1.41\,GPa, respectively.  It should be noted that $T_{\rm sc}$ decreases from 0.45\,K to 0.25\,K although $N_{\rm res}$/$N_0$ does decrease from 0.50 to 0.30 at $P=1.2$\,GPa in FM2 and at $P=1.41$\,GPa in FM1. These results demonstrate that the presence of the RDOS in the superconducting state is not due to the impurity effect but intrinsic in origin. Although some impurity and/or imperfection based effects, if any, are not completely ruled out,  we state that the observation of the highest $T_{\rm sc}=0.75$\,K ensures that the present sample is one of the best quality samples reported thus far.
\\
\indent
First, we address whether or not the RDOS is associated with a self-induced vortex state in the SC~+~FM uniformly coexisting  state. By assuming the Abrikosov triangular vortex lattice, a coherence length $\xi\sim$ 130\,\AA\ \cite{Tateiwa1}, and an internal magnetic field $H= 0.125$\,T \cite{MagneticField}, we consider that only 3\,\% of $N_{\rm res}/N_{0}$ arises from the normal state inside the self-induced vortex core in the SC\,+\,FM state, which does not agree with the experimental result.
Alternatively, in another promising scenario that explains the RDOS, we consider a nonunitary spin-triplet pairing model \cite{Ohmi}. In this model, the superconducting energy gap opens only in the up-spin band parallel to the magnetization of FM phases, but not in the down-spin band that remains gapless.  
We begin by assigning possible nuclear relaxation processes of Ge-NQR $T_1$ in the ferromagnetic states. One of them is caused by the transversal component of fluctuations of internal magnetic fields at the Ge sites originating from intraband and interband transitions across the Fermi level at each up-spin band and down-spin band. Another one is caused by only the interband spin-flip transition across the Fermi level between the up-spin band and down-spin band. By considering these relaxation processes, $1/T_1T$ in the FM state is expressed as  

\[\displaystyle \frac{1}{T_1T}\propto 2t_2(T)\cos^2\theta+[t_1(T)+2t_2(T)+t_3(T)]\sin^2\theta,\]
\[\displaystyle t_1(T)=\frac{1}{k_{\rm B}T}\int^{\infty}_0 dE N^2_{\downarrow}(E)f(E)[1-f(E)],\]
\[\hspace{1.0cm}\displaystyle t_2(T)=\frac{1}{k_{\rm B}T}\int^{\infty}_0 dE N_{\uparrow}(E) N_{\downarrow}(E)f(E)[1-f(E)],\] 
\[\displaystyle t_3(T)=\frac{1}{k_{\rm B}T}\int^{\infty}_0 dE N^2_{\uparrow}(E)f(E)[1-f(E)],\]\\
where $t_1(T)$, $t_2(T)$, and $t_3(T)$ indicate the former contributions and $t_2(T)$ represents the latter contribution which is only possible for $\theta=0$. When the energy dependence of the DOS is neglected near the Fermi level, all contributions  of $t_1=N_{0\downarrow}^2$, $t_2=N_{0\uparrow}N_{0\downarrow}$, and $t_3=N_{0\uparrow}^2$ are independent of temperature. Here, $N_{0\uparrow}$ and $N_{0\downarrow}$ are the DOS at the up-spin and the down-spin bands at the Fermi level in the normal FM state, respectively. $\theta$ is an angle between the quantization axis of the $^{73}$Ge-nuclear-quadrupole Hamiltonian and that of $H_{\rm int}$ at the Ge site in the FM state, which is estimated as $\theta\sim\pi/3$ from the analysis of the NQR spectra in the FM state, and $f(E)$ is a Fermi distribution function. In the ferromagnetic spin-pairing model, the line-node gap of $\Delta(\phi)=\Delta_0\cos\phi$ is assumed only for the density of states $N_{{\rm s}\uparrow}(E)$ at the up-spin band, but not for $N_{{\rm s}\downarrow}(E)$ at the down-spin band. It should be noted that if $\theta=0$, $t_2(T)$ should behave as $1/T_1\propto T^2$ well below $T_{\rm sc}$. In the present case, because of $\theta\sim\pi/3$, the gapless term $t_1$ gives rise to the RDOS at the Fermi level in the superconducting state, as shown in Fig.~4(b). 
In fact the experimental results are actually in good agreement with this theoretical model, as indicated by the solid lines in Figs.~4(a) and 4(b). Therefore, the SC energy gap $\Delta$ and $N_{0\uparrow}$/$N_{0}$ are estimated as $2\Delta/k_{\rm B}T_{\rm sc}\sim 3.7$, 3.8, 4.0, and 3.7 with $N_{0\uparrow}$/$N_{0}$ = 0.57, 0.57, 0.82, and 0.80 at $P$ = 1.17, 1.2, 1.24, and 1.41\,GPa, respectively.  Here, $N_{0}(P)=N_{0\uparrow}(P)+N_{0\downarrow}(P)$.  
\\
\indent
In order to gain further insight into the novel SC state, Fig.~4(b) shows the $T$ dependence of $(1/T_1T)^{1/2}$ related to the DOS at the Fermi level in either the SC or the normal FM state. As shown in Fig.~5(c), the most interesting finding is that $N_{0\uparrow}(P)$ dramatically increases as $P$ increases slightly from $P=1.2$ to 1.24\,GPa across $P_{\rm x}=1.23$\,GPa, accompanying the sudden reduction of $H_{\rm int}$ shown in Fig.~5(b). By contrast, $N_{0\downarrow}(P)$ remains almost constant and the $T$-linear coefficient of the specific heat $\gamma$ gradually increases with $P$ across $P_{\rm x}$ \cite{Tateiwa3}. These results reveal that the  Fermi level in FM2 is located just above a sharp peak in the majority up-spin band, and as $P$ increases across $P_{\rm x}$, it shifts down toward the peak when it enters the FM1 phase. In the ferromagnetic spin-pairing SC state, the large DOS in the up-spin band in FM1 enhances $T_{\rm sc}$, leading to the highest value of $T_{\rm sc}=0.75$\,K, whereas its reduction in FM2 decreases $T_{\rm sc}$, as shown in Figs.~5(a) and 5(c). 
However, it should be noted that as $P$ increases further up to $P=1.41$\,GPa, even though $N_{0\uparrow}(P)$ for FM1 remains rather larger than that for FM2, $T_{\rm sc}=0.25\pm 0.05$\,K for FM1 at $P=1.41$\,GPa becomes lower than $T_{\rm sc}= 0.45$\,K for FM2 at $P= 1.17$\,GPa . In this context, the large increase in $N_{0\uparrow}(P)$ is not always a main factor that increases $T_{\rm sc}$. Rather, the first-order transition from FM2 to FM1 at $P_{\rm x}$ is responsible for the mediation of the up-spin Cooper pairing. The longitudinal FM spin fluctuations along the $a$-axis \cite{Huxley2} softens in energy at the critical end point for the first-order transition at $P_{\rm x}$. Therefore, we suggest that this longitudinal FM fluctuations along the $a$-axis would be a mediator of the ferromagnetic spin-pairing SC where the majority up-spin band in the FM phases is gapped, while the minority down-spin band is not.
%fig4
\begin{figure}[h]
\centering
\includegraphics[width=6.05cm]{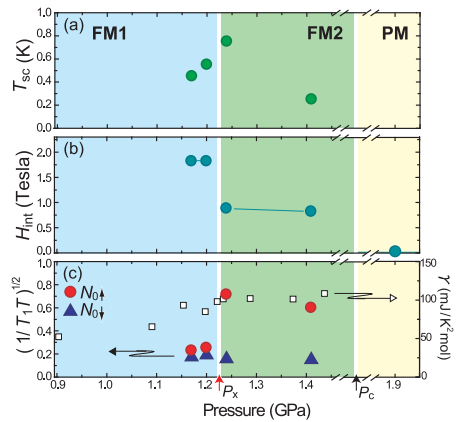}
\caption[]{(color online). \footnotesize Pressure dependence of (a) $T_{\rm sc}$; (b) internal magnetic field, $H_{\rm int}$, at the Ge1 site; and (c) relative $P$ dependence of $N_{0\uparrow}$ (solid circles) and $N_{0\downarrow}$ (solid squares) estimated from the ferromagnetic spin-pairing model on a scale of $(1/T_1T)^{1/2}$ and the $T$-linear coefficient of the specific heat $\gamma$ \cite{Tateiwa3} (open squares) across $P_{\rm x}$ (see text). It should be noted that $N_{0\uparrow}$ dramatically increases when the first-order transition from FM2 at $P=1.2$\,GPa to FM1 at 1.24\,GPa occurs across $P_{\rm x}=1.23$\,GPa.}
\end{figure}
\\
\indent
In another context, it is predicted that $T_{\rm x}$ could be identified with the formation of a simultaneous charge- and spin-density wave (CSDW) induced by the imperfect nesting of the Fermi surface for the up-spin band and hence the superconducting pairing is mediated by CSDW fluctuations around $P_{\rm x}$ \cite{Watanabe}. 
Although the NQR spectrum does not directly evidence an onset of static CSDW states, the remarkable increase in $N_{0\uparrow}$($P$) across $P_{\rm x}$ is relevant to the nesting at the Fermi surface for the up-spin band below $T_{\rm x}$.                       
\\
\indent                    
In conclusion, the $^{73}$Ge-NQR measurements under pressure on well characterized UGe$_2$ have revealed
that the superconducting energy gap opens only with the line-node at the Fermi level in the majority up-spin band, but the down-band remains gapless. It is therefore concluded that ferromagnetic spin-pairing SC occurs in UGe$_2$.  We have also shown that the first-order transition from FM1 to FM2 at $P_{\rm x}=1.23$\,GPa occurs because the Fermi level is located just on the peak in the DOS of the up-spin band.  The ferromagnetic spin-density fluctuations emerging in the vicinity of the critical end point for this first-order transition are considered to be the mediator of the onset of novel SC realized in ferromagnet UGe$_2$.  
\\
\indent
%\section*{Acknowledgment}
We thank H.~Kotegawa, N.~Tateiwa, S.~Watanabe, and K.~Miyake for fruitful discussions and comments. This study was supported by Grant-in-Aid for Creative Scientific Researchi15GS0213); the Ministry of Education, Culture, Sports, Science and Technology (MEXT);
and the 21st Century COE Program (G18) supported by the Japan Society for the Promotion of Science (JSPS).
A.\,H. was financially supported by a Grant-in-Aid for Exploratory Research of MEXT (No.~17654066).
%\appendix
%\vspace{2cm}


\begin{references}
\bibitem{MagneticField}
The internal magnetic field $H$ is derived from $H$ = $(1-n)M$, where $n$ is the demagnetization factor and $M$ is the magnetization per volume. Here, the ferromagnetic moments is given by 1\,$\mu_{\rm B}$/U \cite{Tateiwa2} and the demagnetization factor for this sample geometry (approximately spherical) is $n = 1/3$.
\end{references}
\end{document}